\documentclass[aps,prl,twocolumn,groupedaddress,showpacs,floatfix]{revtex4}
\usepackage{dcolumn}
\usepackage{amsmath}
\usepackage{amssymb}
\usepackage{graphicx}
\usepackage{bm}
\usepackage[T1]{fontenc}
\usepackage{color}
\usepackage{url}
\usepackage[bf]{subfigure}
\usepackage{rotating}

\usepackage{algorithm}
\usepackage{algorithmic}

\begin{document}

\title{What Is There Between Any Two Nodes in a Complex Network?}

\author{Luciano da Fontoura Costa}
\email{luciano@if.sc.usp.br}
\author{Francisco A. Rodrigues}
\affiliation{Instituto de F\'{\i}sica de S\~{a}o Carlos, Universidade de S\~{a}o Paulo, Av. Trabalhador
S\~{a}o Carlense 400, Caixa Postal 369, CEP 13560-970, S\~{a}o Carlos, S\~ao Paulo, Brazil}

\begin{abstract}
This article focuses on the identification of the number of paths with
different lengths between pairs of nodes in complex networks and how,
by providing comprehensive information about the network topology,
such an information can be effectively used for characterization of
theoretical and real-world complex networks, as well as for
identification of communities.
\end{abstract}

\maketitle

\section{Introduction}

A large number of natural and artificial complex systems can be
represented and modeled in terms of networks involving interacting
components. Such interactions can range from signalling between cells
to social contacts (e.g.~~\cite{Amaral04:EPJB}). Indeed, complex
networks theory has been considered in a wide range of applications
including neuronal connections, protein-protein interactions, economy,
Internet communication and social ties~\cite{costa2007aam}, to cite
just a few possibilities. It was thanks to their flexibility and
potential for multidisciplinary applications that complex networks
became so popular and important.

Much of the efforts by networks researchers have been concentrated in
developing tools for characterization, classification, modeling and
simulation. The characterization of network structure is one of the
fundamental steps of complex networks research, because the modeling,
simulation and classification of networks all depend strongly on
accurate descriptions of the respective topology
description~~\cite{Costa_surv:2007, Boccaletti06}. In order to
quantify different topological properties, a large set of network
measurements has been developed~\cite{Costa_surv:2007}. Many of these
features are related to the concept of connectivity between nodes,
taking into account the immediate links between each pair of
nodes. Several of the measurements currently employed in order to
characterize network structure -- such as degree, clustering
coefficient and shortest path length --- are ultimately related to
pairwise interconnectivity~\cite{Costa_surv:2007}.  It is therefore
important to resort to longer range interaction between nodes in order
to achieve more comprehensive description, characterization and
modeling of complex structures.  The average shortest path length (or
geodesic distance) between a pair of nodes corresponds to one of such
measurements~\cite{Watts98:Nature}. Usually, its average value is
obtained considering the shortest distance between every pair of
nodes.  Some works have also considered distance matrices, containing
minimum shortest path lengths, in order to enhance the
characterization~\cite{nicolelis1990scn, Andrade2008ccn,
Bagrow2008pcn, shavitt2007bcc} and identification of
isomorphisms~\cite{Andrade2008ccn, Bagrow2008pcn}. Nevertheless, the
isolated consideration of these measurement results in incomplete
network characterization, since important information about network
structure is not taken into account.  For instance, the alternative
paths between pair of nodes whose lengths are larger than the shortest
path are completely overlooked by more traditional network
analysis. Thus, two networks presenting the same degree and shortest
path distributions, but different alternative paths organization, can
be characterized as being identical, which is clearly
inappropriate. Also, alternative paths can provide additional
information about network resilience, once they generally reinforce
connections, providing alternative routes and maximizing the
flow. More traditional robustness analysis taking into account just
the local connectivity and measurements related to the shortest paths,
such as betweenness centrality~\cite{Motter02:PRE} and
efficiency~\cite{Latora01:PRL}, also do not take into account the
richer interconnectivity structure provided by longer alternative
paths.

The comprehensive characterization of pairwise connectivity clearly
requires more general approaches considering multi-scale interactions
extending from the immediate link to long-range connectivity. The term
multi-scale refers to the varying topological scales which are
progressively taken into account around the nodes. The traditional
approach considers just the first scale ($h=1$), \emph{i.e}.\
immediate neighbor connectivity. In other words, in addition to
immediate-connection measurements and limited long range information
such as the shortest paths, the identification of alternative paths of
any length can enhance the network characterization, providing a more
complete description of network topology. Measurements taking into
account the successive shortest path lengths from a reference node
(concentric neighborhoods) have been proposed in the literature in
terms of hierarchical or concentric representations~\cite{Costa:2004,
Costa_NJP:2007, Costa_JSP:2006, Costa_EPJB:2005,Costa_surv:2007}.

\begin{figure}[t]
\begin{center}
\includegraphics[width=1\linewidth]{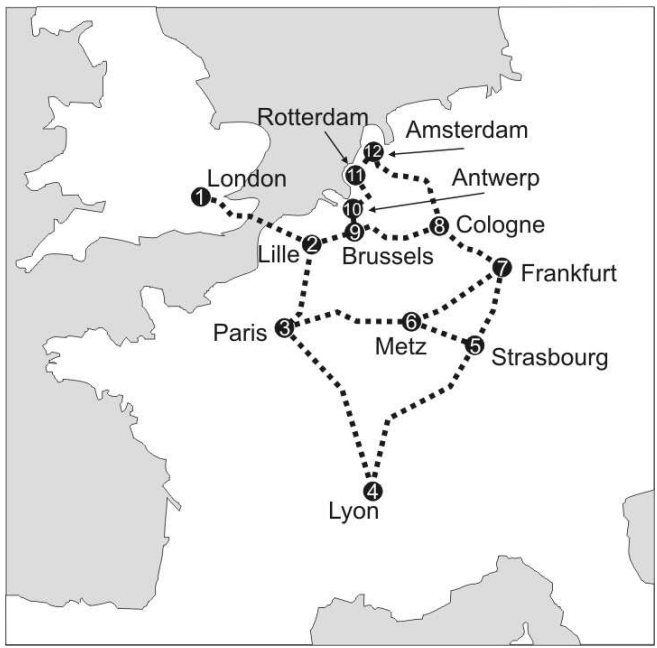} \\
\end{center}
 \caption{The European high-speed rail network connecting the main
 cities of central Europe.  While the traditional network
 characterization in terms of the shortest distance takes into account
 just one path of length three between London and Lyon, the other seven
 alternative paths are overlooked. However, the alternative paths
 are fundamental for network topology and can be associated to
 important dynamics such as traffic jamming and resilience~\cite{Boccaletti06}. }
\label{fig:ex}
\end{figure}

The further generalizations of the concepts of connectivity and
interaction necessarily in order to account for larger portions of the
network, requires the identification of alternative paths between
pairs of nodes, as illustrated in Figure~\ref{fig:ex}.  Let us suppose
we are interested in the pairwise interconnection between London (UK)
and Lyon (France), which is particularly important for those people
who have to travel by express train between those two cities.  If we
consider the shortest path approach, just the path of length three
between those two cities is taken into account, while the other seven
alternative paths are completely overlooked. Nevertheless, such paths
are still fundamental for network communication and resilience. For
instance, if the connection London-Paris-Lyon is blocked in any part
other than from London to Lille, the train or passengers can always
take the alternative routes.  The importance of the identification of
alternative paths in networks has also been substantiated with respect
to the cardiovascular system~\cite{nicolelis1990scn}.

In the current article we report a comprehensive approach to
generalize the concept of pairwise connectivity through the
quantification of the distribution of paths of different lengths
between pairs of nodes. The potential of such a framework is
illustrated with respect to network characterization, with respect to
both theoretical models and real-world networks, as well as community
detection.  Helped by optimal multivariate statistical methods, we
characterize the relationships between the topologies of six distinct
complex networks theoretical models and discuss the achieved
discriminability. In order to illustrate the variation of the
generalized connectivity in real-world networks, we report and discuss
results corresponding to: (i) the US highway network, (ii) the neural
\emph{C. elegans} network~\cite{Watts98:Nature}, (iii) the cat
cortical network~\cite{sporns2004swc} and (iv) a food web of a
broadleaf forest in New Zealand~\cite{Jaarsma1998cfw}. In addition, we
characterize the network modular structure (community) considering
respective generalized connectivity matrices. The projection of the
network vertices considering an optimal multivariate statistical
method resulted in vertices belonging to the same communities being
projected nearby, forming clusters of points.

In next sections, we provide the basic concepts related to network
models, paths between nodes, principal component analysis (PCA) and
network discriminability. An optimal algorithm to find the number of
paths between pair of vertices is also provided. The illustration of
the potential of the proposed methodology with respect to theoretical
and real-world networks, as well as for community identification, are
presented and discussed subsequently.

\section{Basic Concepts and methodology}

\begin{figure}[t]
\begin{center}
\includegraphics[width=1\linewidth]{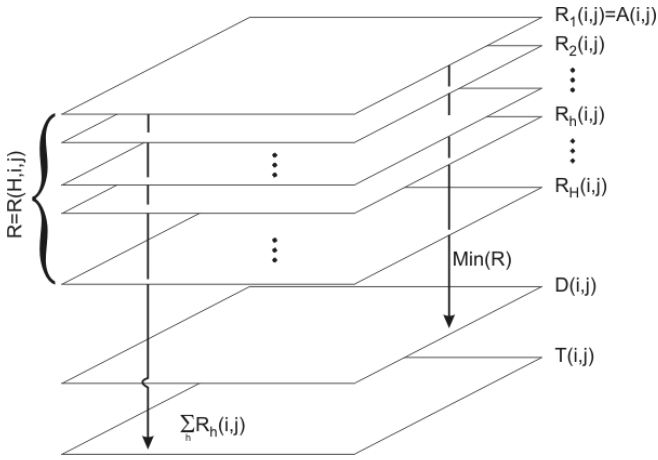} \\
\end{center}
 \caption{Networks can be characterized in terms of the three-dimensional
 matrix $R=R(H,i,j)$, which provides a more comprehensive description of the
 network structure than the traditional adjacency ($A(i,j)=R_1(i,j)$)
 and shortest paths length matrices ($D(i,j) = min(R(H,i,j))$). The
 matrix $T$, which provides the total number of paths between every pair of
 vertices $i$ and $j$, can be obtained by summing the elements of
 the matrices $R_h(i,j)$.}
\label{fig:mat}
\end{figure}

An undirected network can be represented by its adjacency matrix $A$,
whose elements $a_{ij}$ are equal to one whenever there is a
connection between the vertices $i$ and $j$, or equal to zero
otherwise. The number of connections of a given vertex $i$ is called
its degree $k_i$, while the clustering coefficient $cc_i$, is defined
as \emph{i.e}.\ $cc_i = 2n_i/(k_i-1)k_i$, where $n_i$ is the number of
connections between the neighbors of $i$~\cite{Watts98:Nature}. The
number of paths with length $h=1,\ldots,H$ between each pair of nodes
can be expressed in terms of the three-dimensional matrix $R =
R(h,i,j)$ (see Figure~\ref{fig:mat}), so that each matrix $R_h(i,j) =
R(h,i,j)$ gives the total number of paths of length $h = 1, 2, \ldots,
H$ extending from node $j$ to node $i$ (observe that $R_1 = A$). These
matrices will always be symmetric for undirected networks. The set of
matrices $R_h$ therefore conveys comprehensive information about the
generalized connectivity between any pair of nodes, therefore
providing valuable additional information about the network
structure. In addition, the shortest path distance matrix can be
derived from such matrices by taking the minimum value along all
matrices $R$ obtained for all possible $h$ (see
Figure~\ref{fig:mat}). Therefore, the matrices $A$ and $D$ are special
cases of the set of matrices $R_h$. The matrix $T$, which is obtained
by summing the elements along the set $R_h$, gives the number of paths
of lengths $h=1,\ldots,H$ between every pair of vertices. As such,
this matrix quantifies all alternative paths between pair of nodes and
can be used, for instance, in analysis of network resilience.

\begin{figure}[t]
\begin{center}
\subfigure[]{\includegraphics[width=0.48\linewidth]{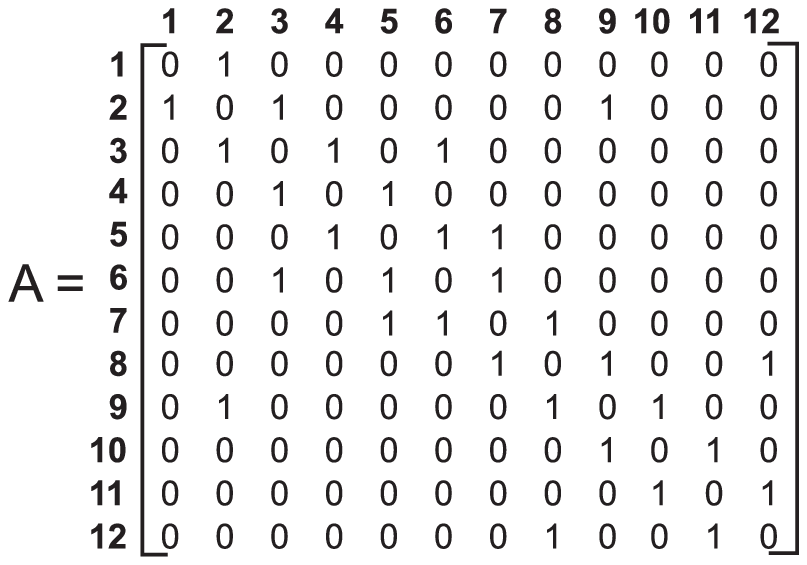}}
\subfigure[]{\includegraphics[width=0.48\linewidth]{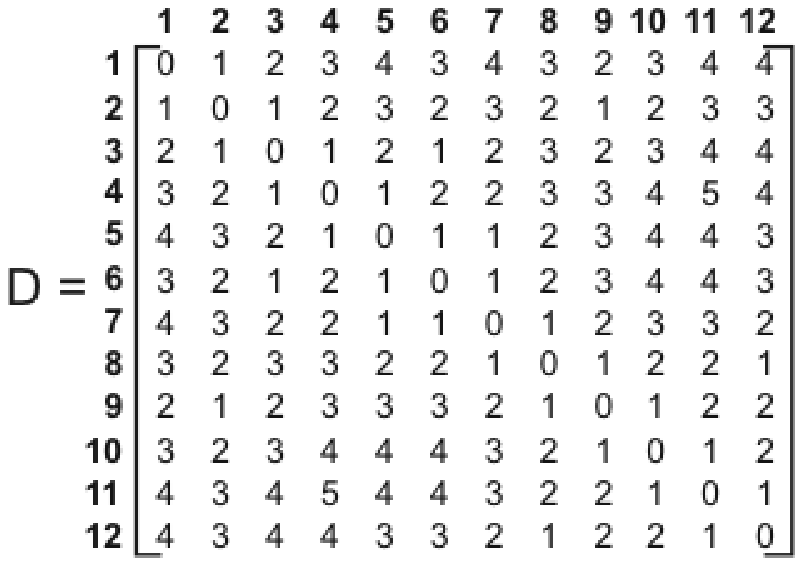}}
\end{center}
 \caption{The adjacency (a) and distance (b) matrices, respective
          to the network in Figure~\ref{fig:ex}.}
~\label{fig:matr-trad}
\end{figure}

\begin{figure}[t]
\begin{center}
\subfigure[]{\includegraphics[width=0.48\linewidth]{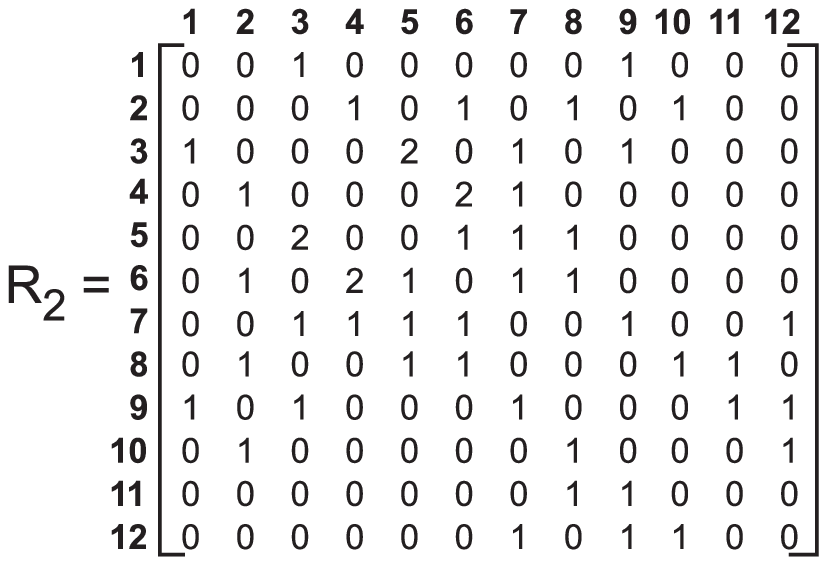}}
\subfigure[]{\includegraphics[width=0.48\linewidth]{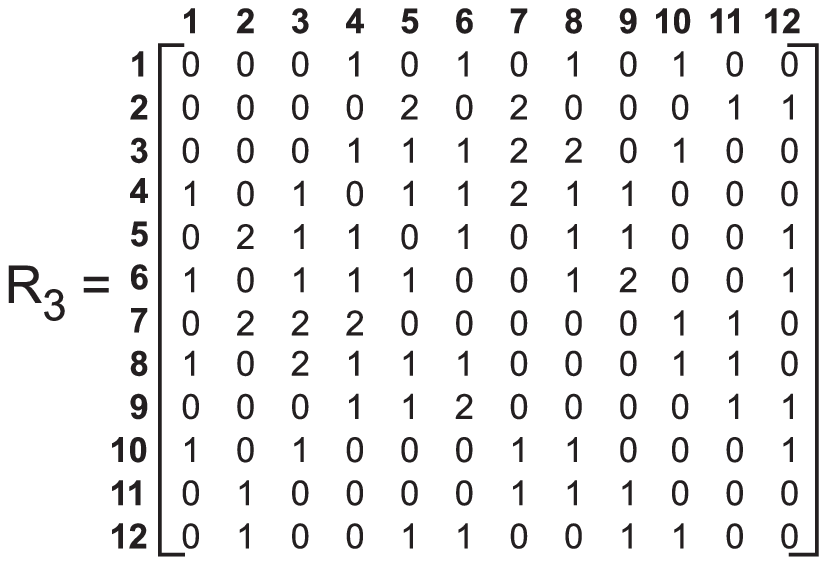}} \\
\subfigure[]{\includegraphics[width=0.48\linewidth]{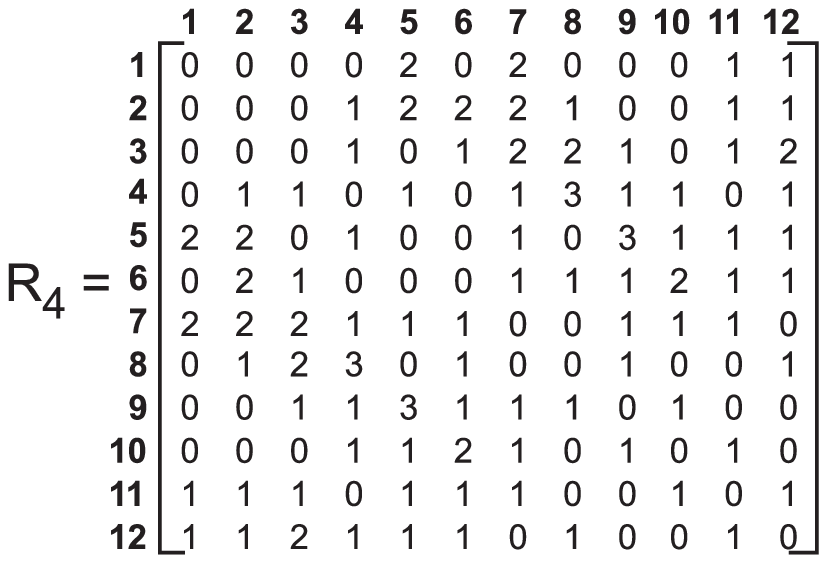}}
\subfigure[]{\includegraphics[width=0.48\linewidth]{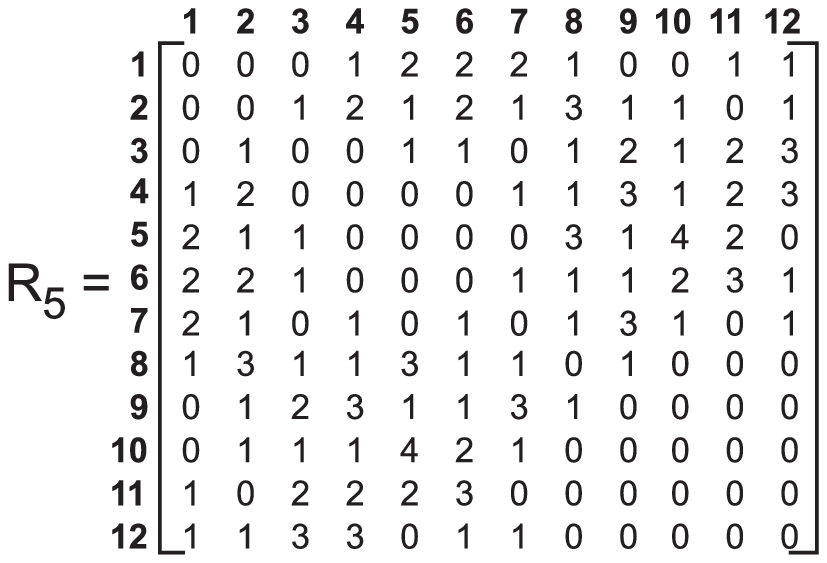}} \\
\end{center}
 \caption{The matrices containing the number of paths of length
          $h=2$ (a), $h=3$ (b), $h=4$ (c) and $h=5$ (d) between
          each pair of nodes in the network in
          Figure~\ref{fig:ex}.}
~\label{fig:matr-path}
\end{figure}

An illustration of the several connectivity approaches that can be
applied in order to characterize the network in Figure~\ref{fig:ex} is
provided in Figures~\ref{fig:matr-trad}
and~\ref{fig:matr-path}. Figures~\ref{fig:matr-trad} (a) and (b) show
the traditionally adopted matrices of adjacency and the shortest path
lengths distances, respectively.  While the adjacency matrix $A$
indicates the immediate connectivity between pairs of nodes, the
shortest path lengths matrix $D$ contains the number of edges along
the shortest paths between each pair of nodes.  On the other hand, the
matrices in Figure~\ref{fig:matr-path} are rarely (if ever) considered
in the literature and express other types of pairwise interactions
between the nodes. In such a figure, the matrices $R_2$, $R_3$, $R_4$
and $R_5$ express the number of distinct paths of lengths $h=2$,
$h=3$, $h=4$ and $h=5$ between each possible pair of nodes in the
network in Figure~\ref{fig:ex}, respectively. Observe that these
matrices make explicit important information which cannot be easily
inferred from any of the two previous matrices, $A$ and $D$. For
instance, while matrix $D$ indicates that there is only one paths of
length four between Strasbourg (France) and Antwerp (Belgium), the
matrix $R_5$ indicates that there are four paths of length five and
matrix $R_4$, a path of length four. Similarly, while the matrix $D$
indicates that there is a single shortest path of length two between
Lyon (France) and Metz (France), the matrix $R_2$, shows two paths of
length two, and the matrix $R_3$, one path of length three, all viable
alternatives in the case of eventual disruption of the shortest
path. The other matrices provide information about even longer
alternative paths, of eventual interest for a tourist who wants to
visit several nearby places.  Therefore, the set of matrices $R$ can
provide valuable additional information about the network structure,
leading to more accurate network characterization and classification.

\subsection{Algorithm for identification of  number of paths}

The Algorithm~\ref{alg_1} allows the identification of all paths
between a reference vertex $i$ and all other nodes in a network.  Such
an optimal algorithm (each node is visited only once) can be applied
to direct and undirected networks. The operations $push(a)$ and
$pop(a)$ place and remove the data $a$ into a stack,
respectively. Though this deterministic algorithm is optimal, it may
require long periods of time depending on the type of network, its
size, average degree, as well as the total number of steps $H$
required.  Stochastic algorithms such as that described
in~\cite{Costa_comp:2007, Costa_longest:2007} can be considered for
estimations in such cases. The execution of such an algorithm from all
vertices on the network yields the set of matrices $R_h$.

\begin{algorithm}[t]
\label{alg} \caption{The general algorithm to obtain the number of paths between each pair of vertices.}
\begin{algorithmic}
\FOR{each vertex $i$}
    \STATE $h=1$;
    \STATE $next$ = one of the non-visited immediate neighbors of $i$;
    \STATE $stack.push$(remainder of non-visited immediate neighbors of $i$, $h$);
    \STATE $path.push$($next$);
    \STATE $R(next,i,h)$ = 1;
     \WHILE{stack not empty or size($path$) > 0 }
        \STATE $curr = next$;
        \STATE $ng$ = number of non-visited immediate neighbors of $curr$;
        \IF{$ng$ > 0}
            \STATE $next$ = one of the non-visited immediate neighbors of $curr$;
            \STATE $stack.push$(remainder of non-visited immediate neighbors of $curr$, $h$);
            \STATE $path.push$($next$);
        \ELSE
            \STATE $next,h$ = $stack.pop$(one node, $h$);
            \STATE $node = -1$;
            \WHILE{$node \neq next$}
                \STATE $node$ = $path.pop()$;
            \ENDWHILE
        \ENDIF
      \STATE $R(next,i,h)$ = $R(next,i,h)$ + 1;\\
    \ENDWHILE
 \ENDFOR
\end{algorithmic}\label{alg_1}
\end{algorithm}

\subsection{Decorrelation of Measurements and Dimensionality Reduction}
~\label{Sec:PCA}

Because of the relatively high dimensionality of the path
measurements, especially as a consequence of their parameterization
with $h$, as well as the already observed correlations along $h$, it
becomes important to consider means for obtaining effective
projections of the measurements (dimensionality reduction) so as to
visualize the network and vertex separations.  This can be optimally
performed through the method know as principal component analysis
(PCA).

PCA can be defined as the orthogonal projection of the original data
onto a lower dimensional linear space, called the principal subspace,
such that the variance of the projected data is maximized along its
first axes~\cite{hotelling1933acs}.  Indeed, PCA can be understood as
a rotation of the axes of the original variable coordinate system to
new orthogonal axes in order to makes the new axes coincide with the
directions of maximum variation of the original
variables~\cite{Costa:book}.  In practice, a PCA consists initially of
finding the eigenvalues and eigenvectors of the sample covariance
matrix~\cite{bishop2006pra}. So, let each of $Q$ observations
(\emph{e.g.} a node, a pair of nodes, or network), henceforth
represented as $v = \{1, 2, \ldots, Q \}$, be characterized in terms
of $M$ respective features or measurements each, represented in terms
of the feature vector $\vec{f_v}$ (each element $f_v(i)$, $i \in \{1,
2, \ldots, M\}$, of this vector corresponds to one measurement of the
observation $v$).  For instance, we can consider the number of paths
between each vertex $i$ and all other vertices in the network. In this
case, each vertex presents a feature vector with $N$ elements. In
cases where the number of features is large, it is possible to
optimally reduce their dimensionality $M$ by removing the correlations
between them.  This important dimensional reduction transformation can
be easily implemented by using the PCA methodology
(\emph{e.g.}~\cite{Costa:book, Costa_surv:2007}).

Let the covariance between each pair of measurements $i$ and
$j$ be given as
\begin{equation}
  C(i,j) = \frac{1}{Q-1} \sum_{v=1}^{Q} (f_v(i) - \mu_i)(f_v(j) - \mu_j),
\end{equation}
where $\mu_i$ is the average of $f_v(i)$ over the $Q$ observations, \emph{i.e}.
\begin{equation}
  \mu_i = \frac{1}{Q} \sum_{v=1}^{Q} f_v(i).
\end{equation}
The covariance matrix between these measurements is defined as $C = [
C(i,j) ]$, with dimension $M \times M$.  Let the eigenvalues of $C$,
sorted in decreasing order, be represented as $\lambda_i$, $i = 1, 2,
\ldots, M$, with respective eigenvectors $\vec{v_i}$.  By stacking
such eigenvectors, it is possible to obtain the matrix
\begin{equation} \label{eq:PCA}
  G  =  \left [ \begin{array}{cccc}
   \uparrow  &  \uparrow   &\ldots   & \uparrow\\
   \vec{v_1}  &\vec{v_2}  & \ldots  & \vec{v_m} \\
   \uparrow  &  \uparrow   &\ldots & \uparrow\\
   \end{array}   \right],
\end{equation}
which defines the stochastic linear transformation known as
Karhunen-Lo\`eve transform~\cite{Costa:book, Costa_surv:2007}. Now,
the new feature vectors can be obtained from the original measurement
vectors $\vec{f}$ by making
\begin{equation}
  \vec{g} = G \vec{f}.
\end{equation}
The variances of the new measurements in $\vec{g}$ are provided by the
respective eigenvalues.  In case the measurements are correlated, most
of their variances will be concentrated along the first elements of
$\vec{g}$, which is guaranteed by the fact that the PCA completely
decorrelates the original measurements.  Indeed, the PCA is optimal
with respect to concentrating the variation along the first axes.
Therefore, it is possible to reduce the dimensionality of the features
vectors by disregarding in the matrix in Equation~\ref{eq:PCA} all
eigenvectors associated to eigenvalues smaller than a given threshold,
or by taking only the $R$ first eigenvectors.  The resulting
variables, which are fully uncorrelated linear combinations of the
original measurements, concentrate the variance of the overall data
and therefore represent a particularly meaningful characterization of
the distribution of the original observations.

\section{Characterization of Theoretical Network Models}

Six different types of theoretical network models are considered in
this article. The Erd\H{o}s-R\'enyi (ER) random
graphs~\cite{Erdos-Renyi59} are obtained by connecting $N$ initially
isolated nodes with constant probability $p$. The traditional
preferential attachment rule~\cite{Barabasi1999} is used to obtain the
scale-free Barab\'asi-Albert (BA) networks. Such a model is a
particular case of the Krapivsky \emph{et al.}~\cite{Krapivsky00}
complex network model, which considers a non-linear preferential
attachment rule to establish connections during network growth --- the
probability of connection is defined as $\mathcal{P}_{i\rightarrow j}
= k_j^{\alpha}/\sum_u k_u^{\alpha}$, where $\alpha$ is the non-linear
exponent. Observe that $\alpha = 1$ yields the BA model. In order to
obtain the Watts-Strogatz small-world model (WS), each connection in a
linear lattice is rewired with probability $p$.  Geographical networks
(GN) are obtained by starting with $N$ nodes distributed uniformly
along a three-dimensional space and connecting them according to
distance, \emph{i.e}.\ the probability to connect two vertices $i$ and
$j$ is given by $P_{ij}=\lambda exp(-\lambda d_{ij})$, where $\lambda$
is a parameter to adjust the network degree and $d_{ij}$ is the
Euclidean distance between $i$ and $j$. Such a model was introduced by
Waxman to model the Internet topology~\cite{waxman1988rmc}. Knitted
networks (KT)~\cite{Costa_comp:2007} can be obtained by generating
random sequences of nodes and connecting them sequentially (without
repetition). The number of generated sequences depends on the network
average connectivity. This network is particularly regular with
respect to several of its topological and dynamical
properties~\cite{Costa_comp:2007, Costa_longest:2007}. In the current
work, all these networks are grown with parameter sets so as to have
the same number $N$ of nodes and approximately the same average
degree.

In order to visualize the network distribution and separation
(discriminability), the set of networks can be projected into a 3D
space of decorrelated measurements. In the current work, we take into
account as original measurements the averages and standard deviations
of each matrix $R_h$. In this case, if we consider a maximum of $H$
distances, we have a set of $2H$ measurements, and each network is
represented by a feature vector $\vec{v} = \{
\mu_1, \sigma_1,
\mu_2, \sigma_2, \ldots, \mu_H, \sigma_H \}$,
where $\mu_h$ and $\sigma_h$ are the average and standard deviation of
the elements in the matrix $R_h$, respectively.  The network
projections obtained by the PCA reflect the network similarities in
terms of their respective feature vectors. Indeed, models that are
mapper nearby in the projected space tend to present similar
topologies.

\section{Community detection}\label{Sec:comm}

Vertices belonging to the same community tend to present similar
patterns of generalized connectivity, i.e. distributions of the number
of paths of varying lengths. Since the generalized distance matrices
provide comprehensive information about the distribution of paths
between nodes, it can be considered for community detection. Thus,
each vertex of a given network is represented by the feature vector
$\vec{x_i}$ corresponding to the respective row (or column, as the
distance matrices are symmetric) $i$ in the matrix $R_h$. Therefore,
each element $j$ of such vector represents the number of paths of
length $h$ between $i$ and $j$. The visualization of the vertex
distribution can then be obtained by PCA projecting the feature
vectors into the three-dimensional space. Thus, the vertices
presenting similar set of attributes tend to be projected nearby,
giving rise to clusters of points. Each of these clusters indicates a
possible community in the original network.

\section{Results and Discussion}

\begin{figure}[t]
\begin{center}
\includegraphics[width=0.9\linewidth]{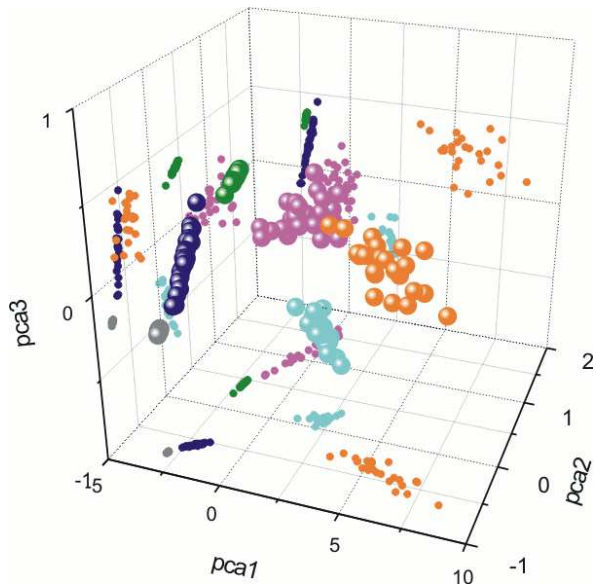} \\
\end{center}
 \caption{The projection of the networks generated by the ER (blue),
 WS (green), BA (orange), GN (magenta), KT (gray), and NL (cyan)
 network models in the 3-dimensional space.}
\label{fig:models}
\end{figure}

Our first experimental investigation concentrates in the
characterization and discrimination between the topologies of six
different complex networks theoretical models, namely: (i) the random
graphs of Erd\H{o}s and R\'enyi (ER), (ii) the small-world network
model of Watts and Strogatz (WS), (iii) the geographical model of
Waxman (GN), (iv) the scale-free model of Barab\'{a}si and Albert
(BA), (v) the non-preferential attachment model of Krapivsky \emph{et
al.} (NL) and (vi) the knitted network model of Costa (KT). We
computed the averages and standard deviations of the matrices $R_h$
for $h=1,\ldots,6$, for each network model realization. In this way,
each generated network is represented in terms of a vector with $12$
elements, \emph{i.e}.\ the network $n$ is represented by the
respective vector $\vec{v}_n = \{ \mu_1, \sigma_1, \mu_2, \sigma_2,
\ldots, \mu_6, \sigma_6 \}$, where $\mu_h$ and $\sigma_h$ stand for the average and
standard deviation of the values in the matrix $R_h$. We generated 25
network realizations for each model and, after
standardization\footnote{ The standardization of a random variable
consists of subtracting its respective average and dividing by the
standard deviation.  The resulting transformed random variable
necessarily has zero mean and unitary standard deviation~\cite{
Costa:book}.}, we projected those networks into the 3D space by
applying the PCA methodology.  As we can see in
Figure~\ref{fig:models}, each of the respective types of networks
generated by these models is represented by independent clusters of
points (sharing similar topological properties), which indicates a
clear separation between each network theoretical model.  While the
networks generated by preferential attachment rule (BA and NL, orange
and cyan points) are organized at the right side of the projection,
the most regular models (KT and ER, gray and blue points) are placed
at the left side.  Indeed, the scale-free networks tend to present
greater variability of the number of paths than the more homogeneous
models, once the presence of hubs tends to increase the number of
paths between pairs of vertices and therefore generates a highly
inhomogeneous path length distribution. In addition, the network
models that generate networks with more regular structure tend to
present the smallest cloud dispersions (KT and WS, gray and green
points). In this way, by providing accurate discriminability between
different models, the generalized connectivity approach presents good
potential for enhancing network characterization and classification.

In the case of the real-world networks, we applied our analysis to:
(i) the US highway network, (ii) the neural \emph{C. elegans}
network~\cite{Watts98:Nature}, (iii) the cat cortical
network~\cite{sporns2004swc} and (iv) a food web of a broadleaf forest
in New Zealand~\cite{Jaarsma1998cfw}. Details about these networks are
given in Table~\ref{Tab:nets}. Since these networks present different
number of vertices and connections we cannot compare them directly --
note that the number of paths for the cortical network is higher than
for the other networks, which is a direct consequence of its higher
average node degree. In this way, we considered the z-score in order
to characterize the distribution of paths, which is calculated
by~\cite{Larsen1981ims}
\begin{equation}
Z_h = \frac{\mu_h-\mu_{random}}{\sigma_{random}},
\end{equation}
where $\mu_h$ is the average number of paths of length $h$ in the real
network, and $\mu_{random}$ and $\sigma_{random}$ are the average and
standard deviation of the number of paths in the respective randomized
network ensemble, which were generated by the configuration model and
present the same degree distribution as the respective real-world
network~\cite{Bender1978anl}. The obtained results for the four
network are presented in Table~\ref{Tab:nets}. It is interesting to
note that just the neural network of the nematode \emph{C. elegans},
which is the only case of a nervous system completely mapped at the
level of neurons and chemical synapses~\cite{Watts98:Nature}, presents
larger number of paths of lengths $h=2,3$ and $4$ than the randomized
counterparts. For $h > 4$, the randomized versions present higher
number of paths.  This suggests that connections of length $2, 3,$ and
$4$ could be more important for allowing proper dynamics in the
\emph{C. elegans} network. The highest difference for $h=3$ suggests
that the evolution of the neuronal organization in this species tended
to favor the alternative connections of length 3, while avoiding
longer range connections. On the other hand, in case of the food web,
the cortical network and the US highway, the z-scores tended to
decrease with $h$, which indicates that such networks tend to present
smaller number of paths of length $h > 2$ than their randomized
versions.  Particularly, since food web tend to present a small number
of trophic levels, there are no paths of length $h > 4$, while the
randomized version can display longer path sizes. Indeed, the small
network diameter is a direct consequence of the energy transmission
between trophic levels ~\cite{power1990efr}.  In the case of the
highway network, the fact that the randomized versions tended to
present larger number of paths than the respective real-world version
is a direct consequence of the fact that the connections in
geographical highway network tend to be constrained by the adjacency
between neighboring localities.

\begin{table*}[t]
\caption{The z-scores and the average number of paths obtained for the
real-world networks.} \label{Tab:nets}
\begin{tabular}{l|c|c|c|c|c|c|c|c|c|c|c|c}
\hline
Network  & $N$ &$\langle k\rangle$ &$Z_2$ & $\langle R_2\rangle$ &$Z_3$ &$\langle
R_3\rangle$ &$Z_4$ &$\langle R_4\rangle$ &$Z_5$ &$\langle R_5\rangle$ &$Z_6$ &$\langle R_6\rangle$\\
\hline
Food web      & 78  & 3.1 & 0.002 &0.05 &-0.087 &0.03 &-0.12 &0       &-0.13  &0     &-0.11  &0\\
Cortical net. & 53  &15.5 &-0.027 &   5 &-0.043 &  85 &-0.06 & 1400   &-0.08  &21800 &-0.11  &331000\\
Neural net.   &297  & 7.9 & 0.026 &0.30 & 0.045 &   3 & 0.01 &   25   &-0.05  &210   &-0.10  &1750\\
US Highway    &284  & 6.0 &     0 &0.02 &-0.048 &   2 &-0.06 &   13   &-0.06  &100    &-0.06  &680\\
\hline
\end{tabular}
\end{table*}

\begin{figure}[t]
\begin{center}
\includegraphics[width=0.9\linewidth]{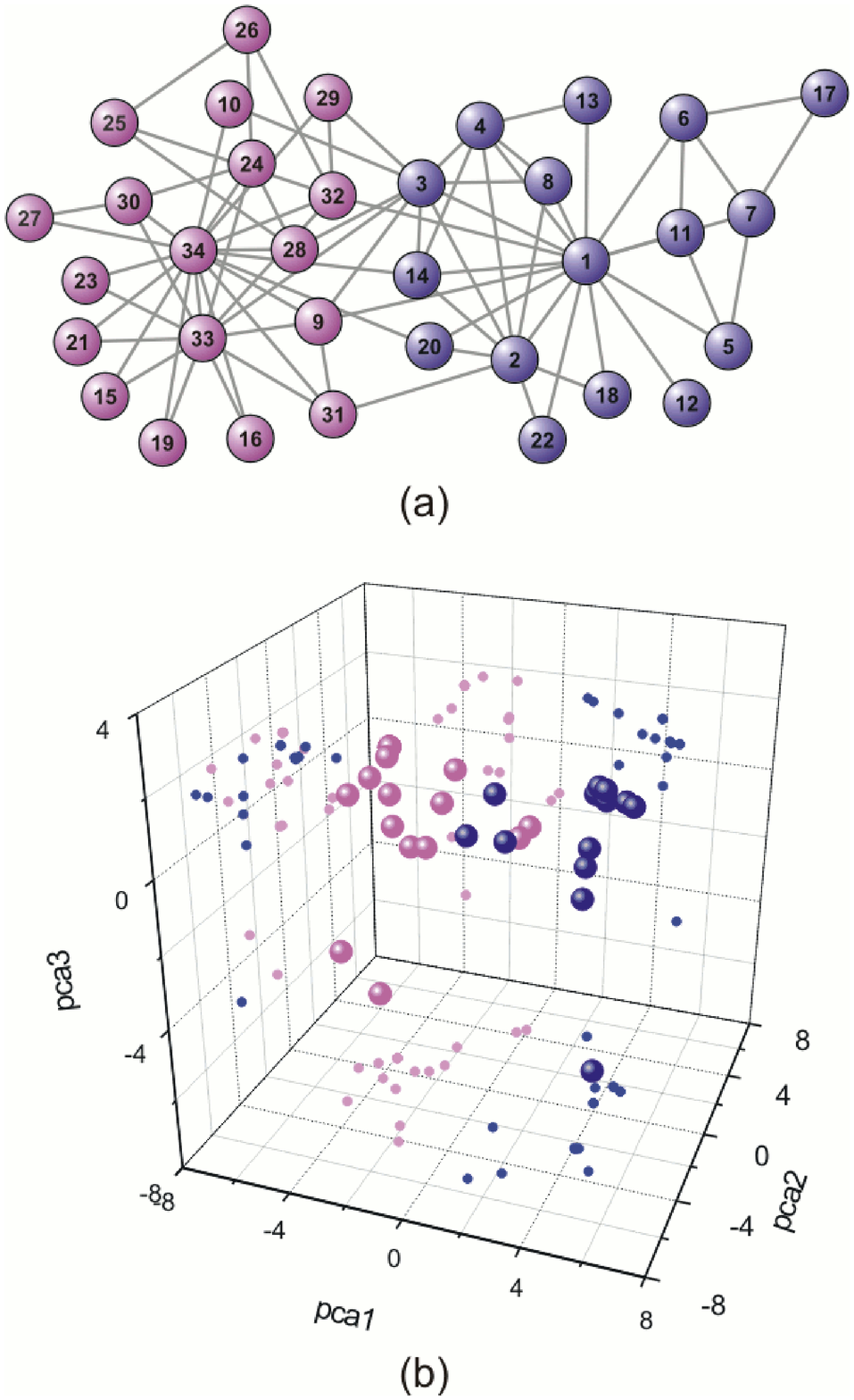} \\
\end{center}
 \caption{The original separation between the two classes of karate
 club member (a), and the projection into the three-dimensional space
 of the generalized matrix $R_2$.} \label{fig:znet}
\end{figure}

\begin{figure}[t]
\begin{center}
\includegraphics[width=0.9\linewidth]{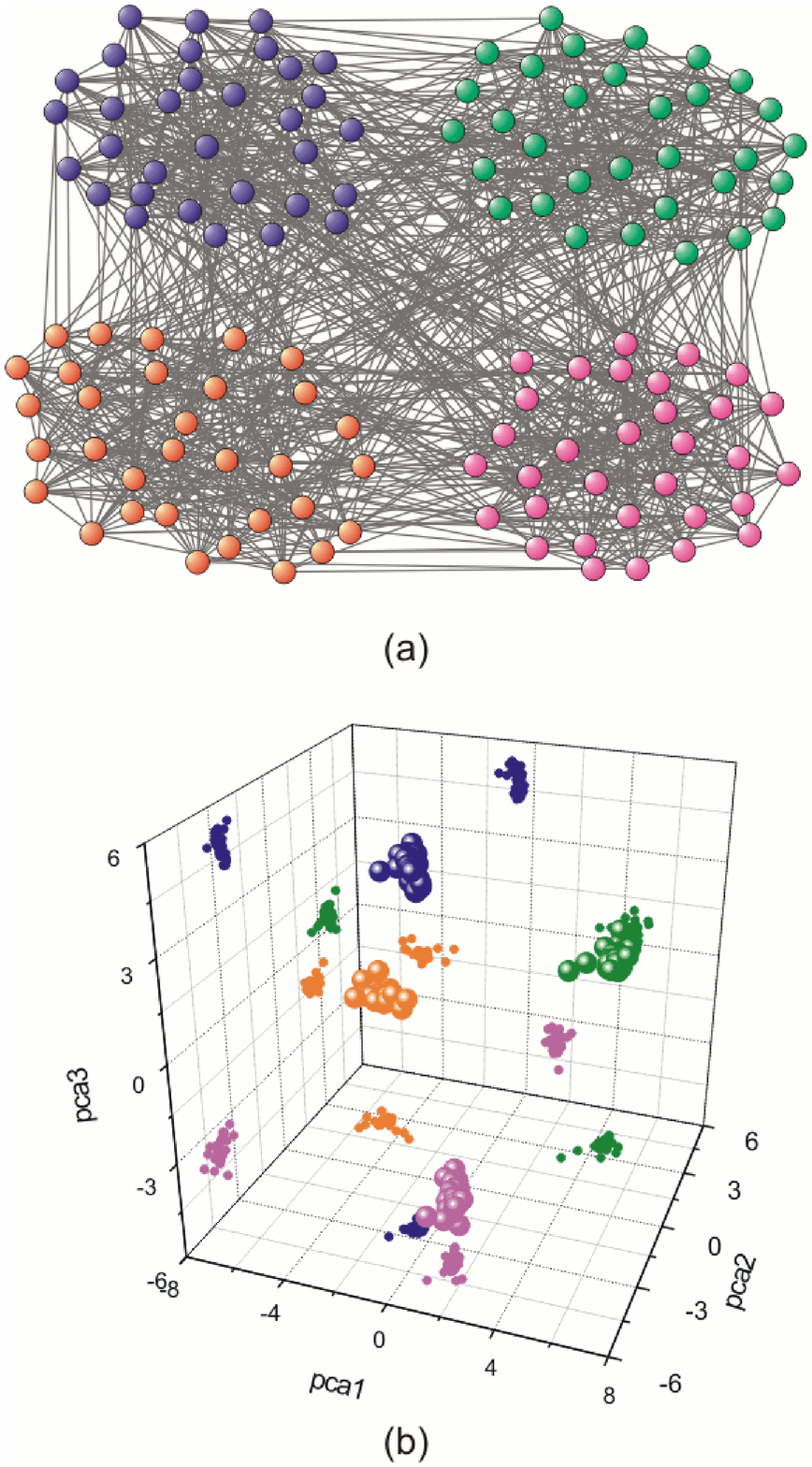} \\
\end{center}
 \caption{(a) The artificial network containing four communities and
 (b) the projection of the respective matrix $R_2$ into the three
 dimensional space considering the PCA methodology.}  \label{fig:com}
\end{figure}

Our final analysis concentrated on the relationship between the
modular network organization and the distribution of the number of
alternative paths. Since vertices in the same community tend to
present similar sets of more strongly connected nodes, the number of
paths between the vertices in the same module tends to be large.  We
applied the proposed methodology described in Section~\ref{Sec:comm}
to the Zachary karate club network and to an artificial modular
network, which have been widely used as tests for community structure
algorithms (\emph{e.g.}\cite{Girvan2002css}). The karate club network
was constructed with the data collected observing $34$ members of a
karate club over a period of $2$ years and considering friendship
between members~\cite{Zachary1977ifm}. On the other hand, the
artificial network was generated as described in~\cite{Girvan2002css},
where a set of $N$ vertices is divided into $c$ communities. Then,
each vertex is connected to $z_{in}$ vertices in the same community,
and $z_{out}$ vertices in the other communities. The connections
between communities are distributed uniformly. In the current work, we
considered $N=128$, $c=4$, $z_{in}=10$ and $z_{out}=6$.  From these
networks, we calculated the respective $R_h$ matrices for $h=1,2$ and
$3$.  After standardization of the feature vectors, we applied the PCA
and obtained the projections presented in Figure~\ref{fig:znet} and
\ref{fig:com} for the karate and the artificial modular networks,
respectively. In case of the Zachary karate club network, the best
identification of the communities was obtained for $h=2$, where the
classification of the vertices into the two clusters corresponds
precisely to the actual division of the club members. The case $h=1$,
which considers the traditional adjacency matrix, does not provide an
accurate separation of the communities into different clusters. For
$h\geq3$, the separation is worse than for $h=2$ because the network
presents a very small average shortest distance ($\ell =
2.3$). Considering the shortest path matrix, the discriminability also
resulted worse than that obtained for $R_2$. In the case of the
artificial modular network, the best separation between the
communities was also obtained for $h=2$, although for $h=3$ and $h=4$
the separation into four respective clusters is still clear. For the
traditional matrices $A$ and $D$, two communities were all joined into
the same cluster, therefore completely undermining the separation. It
is interesting to note that most community finding algorithms cannot
determine the communities perfectly for
$z_{out}=6$~\cite{Boccaletti06}.  Therefore, the consideration of the
alternative paths can provide more information about the network
structure and organization.

\section{Concluding Remarks}

The concept of connectivity underlies great part of the complex
networks research.  However, connectivity has typically been
understood and quantified in terms either of strictly local
measurements, such as the local degree, or by considering shortest
path lengths.  Though more global, the latter feature fails to take
into account alternative pathways between pairs of nodes, which are
extremely important in influencing the topological properties of the
networks. For instance, the presence of more than one path between two
nodes tends to increase the interaction between them and consequently
raises their communication robustness under edge disruption.

In the current paper, we analyzed the generalized network connectivity
with respect to the characterization of six network models and four
real-world networks, as well as for community finding. We showed that
the consideration of the alternative paths between vertices tends to
provide an accurate network topology discriminability, as observed for
the networks generated by the different models. The analysis of
real-world networks suggests that the long range connectivity tend to
be limited in those networks and may be strongly related to network
evolution and organization. In addition, we studied how the
distribution of the number of paths is related to network modular
structure. The obtained results indicate that the proposed approach
particularly promising for community identification.  Indeed, a
possibility for future work would be the improvement of the community
analysis considering clustering methods to separate the cloud of
points obtained in the projection, such as $k-$means or agglomerative
hierarchical clustering~\cite{Costa:book}. In addition, pattern
recognition approaches can be considered in order to quantify the
separation between several types of networks models and therefore
provide complex networks taxonomies. In this case, real-world networks
can be associated to the most likely theoretical model, as described
in~\cite{Costa_surv:2007}.  Studies relating the number of paths with
network dynamics constitute another promising research possibility.

\section*{Acknowledgments}

Luciano da F. Costa thanks CNPq (301303/06-1) and FAPESP (05/00587-5)
for sponsorship.  Francisco Aparecido Rodrigues is grateful to FAPESP
(07/50633-9).

\bibliographystyle{unsrt}
\bibliography{paper}
\end{document}